\documentclass[12pt]{iopart}
\usepackage{iopams}
\usepackage{graphicx}

\begin{document}

\title[P. Chamorro-Posada \etal]{Nonlinear Bloch modes, optical switching and Bragg solitons in tightly coupled micro-ring resonator chains}

\author{P. Chamorro-Posada,$^{1,*}$ P. Martin-Ramos,$^1$ J. S\'anchez-Curto,$^1$ J.C. Garc\'{\i}a-Escart\'{\i}n,$^1$ J.A. Calzada,$^2$ C. Palencia$^3$  and A. Dur\'an$^{3}$}

\address{
$^1$Dpto. de Teor\'{\i}a de la Se\~nal y Comunicaciones e Ingenier\'{\i}a TelemÃ¡tica, Universidad de Valladolid,\\
 ETSI Telecomunicaci\'on, Campus Miguel Delibes s/n, 47011 Valladolid, Spain.\\
$^2$Dpto. de Matem\'atica Aplicada, Universidad de Valladolid, Escuela de Ingenierías Industriales ,\\
 Paseo del Cauce 59, 47011 Valladolid, Spain.\\
$^3$Dpto. de Matem\'atica Aplicada, Universidad de Valladolid,\\
 ETSI Telecomunicaci\'on, Campus Miguel Delibes s/n, 47011 Valladolid, Spain.\\
}

\ead{pedcha@tel.uva.es}

\begin{abstract}

 We study nonlinear wave phenomena in coupled ring resonator optical waveguides in the tight coupling regime.  A discrete model for the system dynamics is put forward and its steady state nonlinear Bloch modes are derived.  The switching behavior of the transmission system is addressed numerically and the results are explained in the light of this analytical result.  We also present a numerical study on the spontaneous generation of Bragg solitons from a continuous-wave input.
\end{abstract}


\maketitle

\section{Introduction}

Transmission media consisting of chains of coupled micro-ring resonators find many applications in Photonics.  They have been proposed for the implementation of optical filters \cite{capmany,chamorro11} or the realisation of fast and slow wave structures \cite{boyd,fraile,chamorro}. One particular benefit of slow-wave optical systems is the associated enhancement of the nonlinear optical response \cite{chen,melloni}.  We study the nonlinear dynamics of a coupled resonator optical waveguide (CROW) built with micro-rings with a Kerr optical response.

{ Similarly to the case of nonlinear Bragg gratings \cite{parini07}, nonlinear CROWs exhibit a wealth of phenomena including multistability \cite{dumeige}, self-pulsation \cite{grigoriev2011,maes2009}, chaos \cite{grigoriev2011}, modulational instability \cite{huang2009} and the generation of isolated gap solitons\cite{christo}.  The dynamics of nonlinear CROWs had been previously addressed using coupled mode theory for short chains \cite{maes2009,grigoriev2011} and discrete nonlinear Schr\"odinger models for long structures \cite{huang2009,christo}.  We use a fully discrete model for chains of arbitrary size, which is a high-order generalisation of the Ikeda equations for the nonlinear ring resonator \cite{ikeda}, where the half-ring length and the associated propagation delay are the natural space and time discretisation intervals. The model equations are introduced in Section 2.  In Section 3 we obtain the nonlinear Bloch modes in the structure.  Numerical surveys on the switching behaviour and the spontaneous generation of a train of localised pulses from a continuous-wave (CW) input are the subject of Sections 4 and 5, respectively.}

\begin{figure}[ht]
\centering
\includegraphics[]{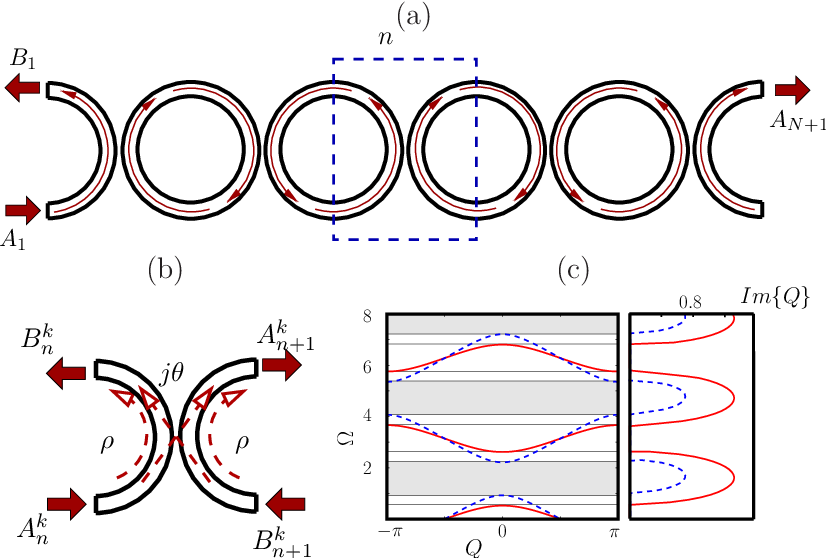}
\caption{(a) Coupled ring resonator optical waveguide. (b) Unit cell. (c) Band structure (first Brillouin zone) for a linear chain for $\theta=0.5$ (solid) and $\theta=0.8$ (dashed).}\label{figu:modelo}\label{figu:bandas}
\end{figure}

\section{Model Equations}

The CROW geometry is displayed in Fig. \ref{figu:modelo}.a. We assume that the evanescent coupling between sections, with a phase jump of $\pi/2$ across the coupled waveguides, is localised at a single point and is described by the parameters $\theta$ and $\rho$, $\theta^2+\rho^2=1$, as illustrated in Fig. \ref{figu:modelo}.b.  Also, we neglect both waveguide and material dispersion, since dispersive effects with origin in the resonator structure itself will dominate the former.  The unit cell defined in Fig.\ref{figu:modelo}.b provides a discretisation scheme for both space and time.  The propagation length across one section is $L=\pi R$, with $R$ the resonator radius, and the sampling period is the corresponding propagation delay $\tau=L/v_g$, where $v_g$ is the group velocity in the optical waveguide sections.  We consider a continuous-wave input with frequency $\omega$ and  will denote with a superscript the time index, where instant $k$ corresponds to $t_k=k \tau$.  Thus, the complex field envelopes of the forward and backward waves in the structure at discrete position $n$ and time $k$ are $A_n^k=A_n(t_k)$ and $B_n^k=B_n(t_k)$, respectively.

{Our discrete evolution equations are a generalisation of the one-ring Ikeda model \cite{ikeda} to an $N-$th order coupled chain  and is obtained by simply keeping track of the signal propagating in the structure.} The signal transmission across the cell length $L$ has associated a linear phase $\exp(-j\Omega)$, with $\Omega=\omega L/v_f$ and a linear loss factor $a=\exp(-\alpha L)$ where we assume a distributed loss with coefficient $\alpha$ and $v_f$ is the phase velocity.  We will set $a=1$ in the lossless case.  The effect of the nonlinear phase will be computed along sections of length $d=L/2$. The nonlinear phase shift is $-\Gamma  |E|^2$, where $\Gamma=\gamma d_{eff}$, $\gamma$ is the nonlinear Kerr coefficient and $d_{eff}=\left(1-\exp\left(-{2}\alpha d\right)\right)/(2\alpha)$ { is an effective propagation length that permits to obtain the exact effect of the nonlinearity by integrating the profile of the variation of the optical intensity $\exp(-2\alpha z)$ along the propagation length $d$ \cite{agrawal}}.  In the lossless case, $d_{eff}=d$. 
The discrete spatio-temporal dynamics are then described as
\begin{eqnarray}
A_{n}^{k+1}&=a\left[j\theta A_{n-1}^k\exp\left(-j\Gamma\left|A_{n-1}^k\right|^2\right)+\right. \nonumber\\
&\left. \rho B_{n}^k\exp\left(-j\Gamma\left|B_{n}^k\right|^2\right)\right]\exp\left(-j\Omega\right)\exp\left(-j\Gamma\left|A_n^{k+1}\right|^2\right)\nonumber\\
B_{n}^{k+1}&=a\left[j\theta B_{n+1}^k\exp\left(-j\Gamma\left|B_{n+1}^k\right|^2\right)+\right. \nonumber\\
&\left. \rho A_{n}^k\exp\left(-j\Gamma\left|A_{n}^k\right|^2\right)\right]\exp\left(-j\Omega\right)\exp\left(-j\Gamma\left|B_n^{k+1}\right|^2\right).\label{eq:modelo}
\end{eqnarray} 
Equations (\ref{eq:modelo}) can be written in time-explicit form substituting
\begin{eqnarray}
\left|A_{n}^{k+1}\right|=a\left|j\theta A_{n-1}^k\exp\left(-j\Gamma\left|A_{n-1}^k\right|^2\right)\right.
\left.+\rho B_{n}^k\exp\left(-j\Gamma\left|B_{n}^k\right|^2\right)\right|\nonumber\\
\left|B_{n}^{k+1}\right|=a\left|j\theta B_{n+1}^k\exp\left(-j\Gamma\left|B_{n+1}^k\right|^2\right)\right.
\left.+\rho A_{n}^k\exp\left(-j\Gamma\left|A_{n}^k\right|^2\right)\right|
\end{eqnarray}
in the corresponding exponential terms in Eqs. (\ref{eq:modelo}).  In this work, we will restrict our analysis to the lossless case $a=1$.  { Refering to Fig. \ref{figu:bandas}.b, Equations (\ref{eq:modelo}) are obtained as tracking the signal $A_{n}^{k+1}$ as a function of $A_{n-1}^{k}$ and $B_{n}^{k+1}$, and  $B_{n}^{k+1}$ as a function of $A_{n}^{k}$ and $B_{n+1}^{k}$}.

The numerical solutions are obtained with a computer code designed for more general propagation conditions \cite{chamorro,fraile} which allows for signal fluctuations at a scale smaller than $\tau$, but comprises any result from the model equations (\ref{eq:modelo}) whenever the dynamics are kept within the time scale $\tau$. {  The computation follows the same scheme employed in the derivation of the theoretical model, but a finer discretisation within the ring sections is employed.  The numerical scheme has been assessed comparing the computational results with the exact solutions of Eqs. (\ref{eq:modelo}), as described below.}

  In our numerical experiments, we { will consider two types of boundary conditions.  In the first one, we impose} a CW input of frequency $\Omega$, $A_1^k=1$, and null input signal at the opposite end of the structure, $B_{N+1}^k=0$.  {These boundary conditions are consistent with previous analyses \cite{maes2009}  and they are a direct representation of the most common implementation of the micro-ring CROW analysed in this work.  The physical layout of such a prototypical structure is depicted in Fig. \ref{figu:modos_ev}.a.  If we define the input and output signal planes as the dashed-dotted lines displayed in the figure, we obtain a direct correspondence with our scheme of Fig. \ref{figu:modelo}.a. } 

{ 
The second type of boundary conditions aims to simulate a reflectionless structure and allows a direct comparison of the numerical results with those of the theory presented here.  To do so, a short spatial region in a very long chain is observed for a time large enough to set up a stationary state and sufficiently small for assuring that no effect originated at the distant end of the structure is detected.  

For both types of boundary conditions, the initial transients can be smoothed out if a slowly growing leading pulse edge is employed instead of a step-like signal.
}

\section{Nonlinear Bloch modes} 

The steady state response for a 1-ring CROW has been shown to exhibit transmission bistability {\cite{fraile91}} and multi-stability has been reported for the $N$-ring system \cite{dumeige}.  Even for an isolated resonator, the micro-ring will display a Ikeda-type instability \cite{ikeda} and behave as a chaotic oscillator \cite{prol}.  The CROW can be described as a coupled array of such oscillators and, not surprisingly, the numerical simulations often show a chaotic response when the system is driven into highly nonlinear regimes.  Therefore, we will restrict our analyses to tightly coupled micro-ring chains, $\theta \to 1$, where signal transmission across resonator stages is large and, thus, stable (nonlinear) propagation effects can be efficiently addressed.  Regarding the transverse instabilities of the type described in \cite{newell}, we have already assumed that the sizes of the micro-rings are too small for these to be of any relevance when neglecting material and waveguide dispersion. 

Signal propagation in the CROW can be analysed drawing a discrete analogue of the Floquet-Bloch theory for continuous periodic systems \cite{kivshar}.  For an infinite chain, we consider steady-state solutions of the type
\begin{equation}
A_n=A\exp\left(-jnQ\right),\,\,B_n=B\exp\left(-jnQ\right),\label{eq:bloch}
\end{equation}
characterised by the mode wave-number $Q$.  For the linear case, we set $\Gamma=0$ in Eqs. (\ref{eq:modelo}) and obtain the the well-known \cite{melloni} linear dispersion relation
\begin{equation}
\cos Q=\sin \Omega/\theta, \label{eq:dispL}
\end{equation}
which provides a relation between $Q$ and $\Omega$ that is periodic in both variables.  Fig. \ref{figu:bandas}.c shows the first Brillouin zone ($Q\in [-\pi,\pi]$) for two different values of $\theta$.  The dispersion relation gives linear wave-numbers $Q$ which become complex in a range of $\Omega$ values defined by the condition $\left|\sin\Omega\right|/\theta>1$.  These regions, where signal propagation is forbidden, define the bands that are highlighted in Fig. \ref{figu:bandas}.c.  We will focus our analysis on the second forbidden band in Fig. \ref{figu:bandas} at the edge of the first Brillouin zone $Q=\pi$.  The band-gap spans the interval $[\Omega_1,\Omega_2]$ with $\Omega_1=\pi+ \arcsin(\theta)$ and $\Omega_2=2\pi-\arcsin(\theta)$. Fig. \ref{figu:bandas} also illustrates the narrowing of the bands as $\theta$ increases, which results in a very narrow gap when we approach the tightly-coupled limit $\theta \to 1$.

\begin{figure}[ht]
\centering
\begin{tabular}{cc}
(a)&(b)\\
\includegraphics[width=8cm]{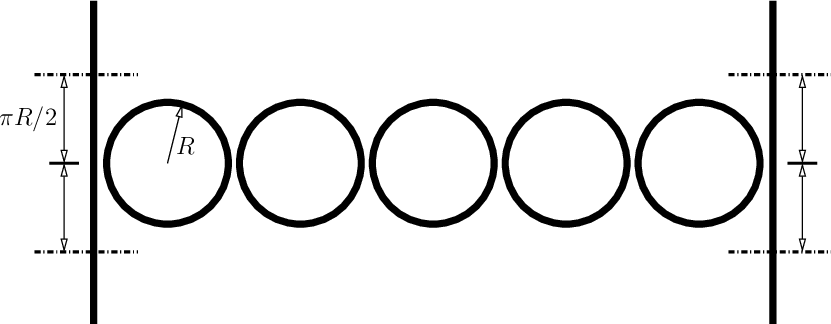}&\includegraphics[width=8cm]{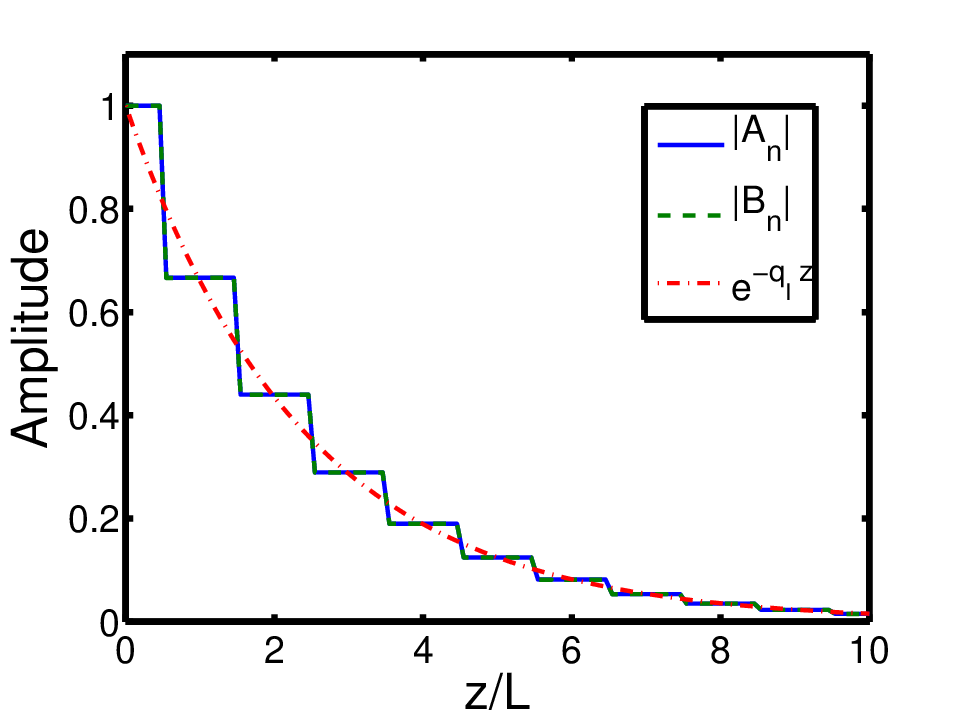}\\
(c)&(d)\\
\includegraphics[width=8cm]{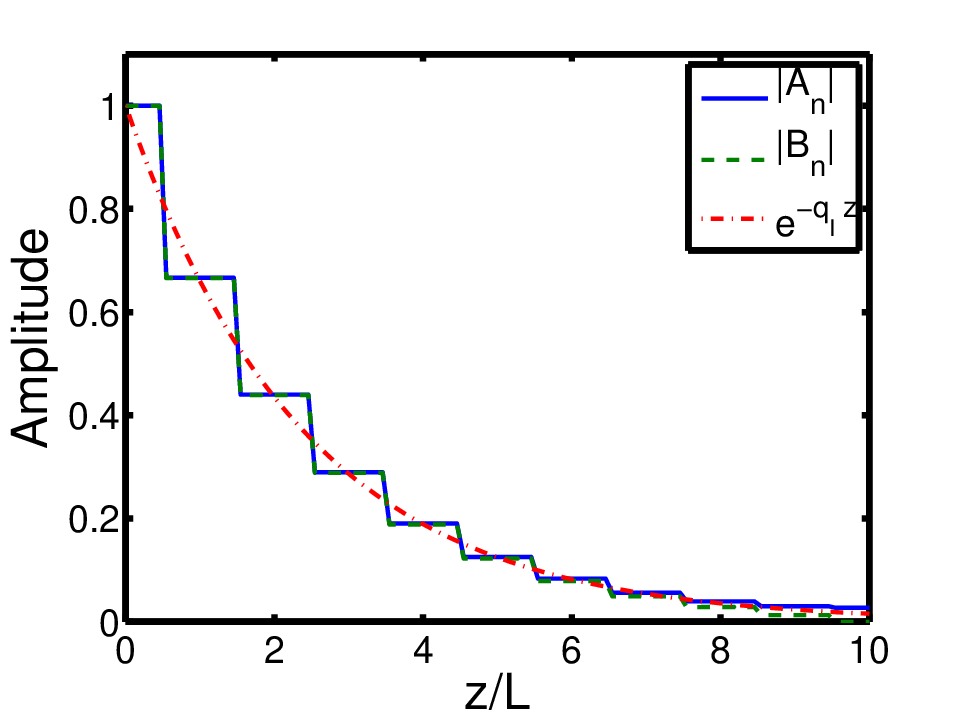}&\includegraphics[width=8cm]{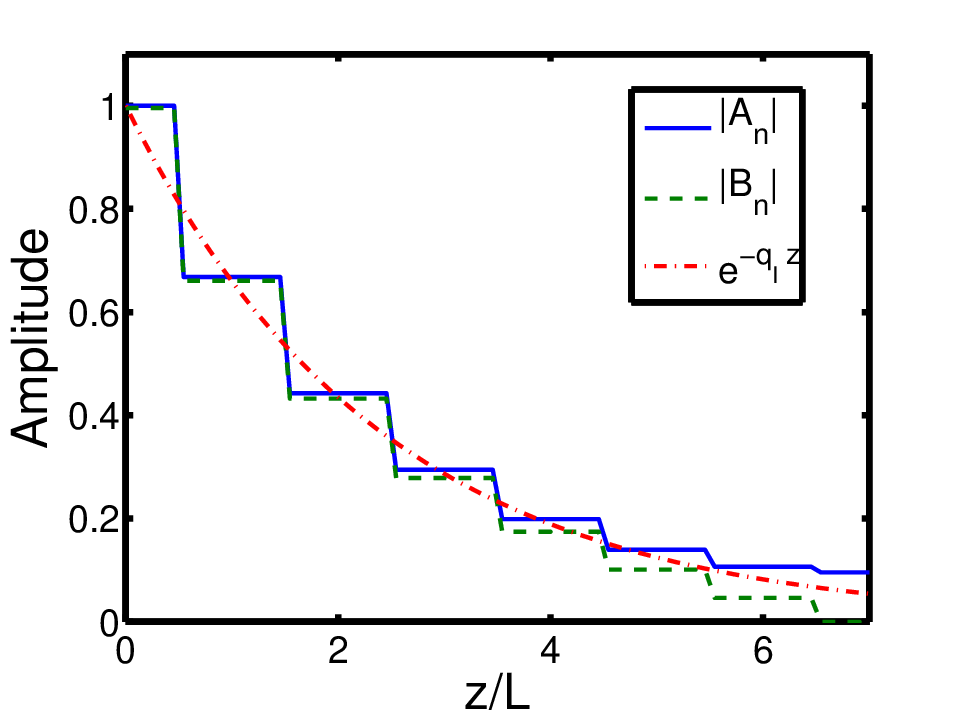}
\end{tabular}
\caption{(a) Schematic of the typical layout of a CROW implementation. (b) to (d) $A$ and $B$ components of the field inside a linear ring chain as obtained from the numerical method and predicted by the theory (dashed-dotted line) for a chain of infinite length: case (b) corresponds to simulated non-reflected boundaries whereas (c) and (d) are simulations for finite-size devices with $N=10$ and $N=6$, respectively.  Parameters are $\Omega=4.9052$ and  $\theta=0.9$.}\label{figu:modos_ev}
\end{figure}

In the full nonlinear case, we define $P=|A|^2+|B|^2$ and the amplitude ratio $f=B/A$.  The mode amplitudes can be expressed as $|A|=\sqrt{P/(1+|f|^2)}$ and $|B|=|f|\sqrt{P/(1+|f|^2)}$.  
Substitution of (\ref{eq:bloch}) into (\ref{eq:modelo}) gives the nonlinear dispersion relation
\begin{equation}
\cos\left(Q+\Delta Q \right)=\sin\left(\Omega+\Gamma P\right)/\theta,\label{eq:dispersion_nl}
\end{equation}
where 

\begin{equation}
\Delta Q = \Gamma P (|f|^2-1)/(|f|^2+1)  \label{eq:def_DQ}
\end{equation}
{ is always a real quantity.}

{ The forbidden band for nonlinear propagation is defined by the relation 
\begin{equation}
\left|\sin\left(\Omega+\Gamma P\right)\right|>\theta \label{eq:condicion}
\end{equation}
which makes the argument of the cosine in Eq. (\ref{eq:dispersion_nl}) purely imaginary (up to, possibly, an additive factor equal to $\pi$).  Therefore, for non-propagative nonlinear Bloch modes, we have two simultaneous conditions on the value of $\Delta Q$ (that it is real and purely imaginary) that are compatible only for $\Delta Q=0$.  In turn, this implies, according to (\ref{eq:def_DQ}), that $\left|f\right|=1$. In this case,  $Q=q_R-iq_I$ with 
\begin{equation}
q_I=\cosh^{-1}\left(\left|\sin(\Omega+\Gamma P)\right|/\theta\right).\label{eq:ate}
\end{equation}

If Eq. (\ref{eq:condicion}) is not fulfilled, we have propagative solutions with real $Q$.  In this case, if we assume a stationary solution of the propagative mode in Equations (\ref{eq:modelo}) we obtain two equivalent conditions that implicitly define $|f|^2$.  For the first of the two equations such relation is 
\begin{equation}
\left|f\right|^2=\left(1+\theta^2-2\theta \sin\left(\Omega-Q+\frac{2\Gamma P}{1+\left|f\right|^2}\right)\right)\left(1-\theta^2\right)^{-1}.\label{eq:f}
\end{equation}
}

\begin{figure}[ht]
\centering
\begin{tabular}{cc}
(a)&(b)\\
\includegraphics[width=8cm]{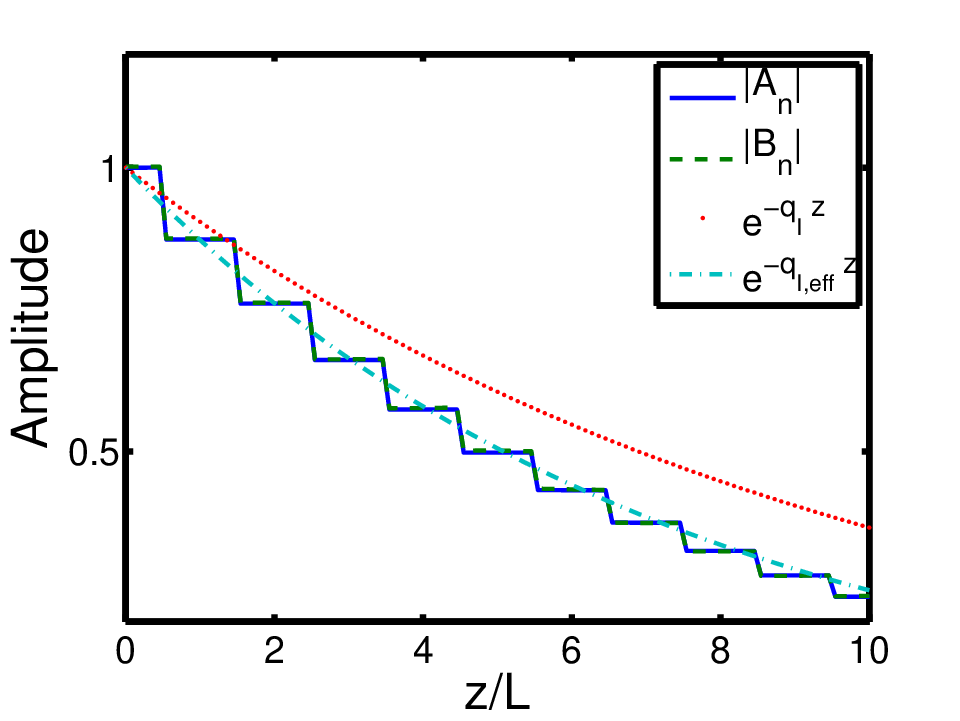}&\includegraphics[width=8cm]{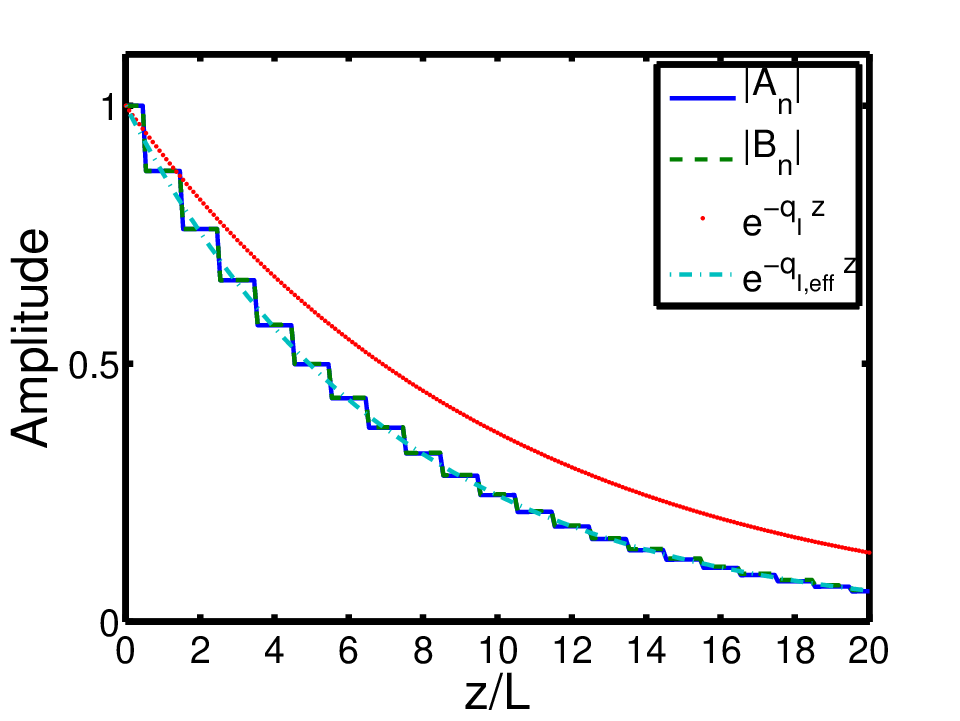}
\end{tabular}

\caption{Nonlinear non-propagative stationary solutions $\theta=0.99$, $\Gamma=0.05$ and $\Omega=4.7124$. Solid line is the $A$ field and dashed the $B$ field. (a) $N=10$ rings,  $g_{eff}=0.391$, and (b) $N=20$ rings with $g_{eff}=0.203$ .}\label{figu:fit}
\end{figure}

\begin{figure}[ht]
\centering
\begin{tabular}{cc}
(a)&(b)\\
\includegraphics[width=8cm]{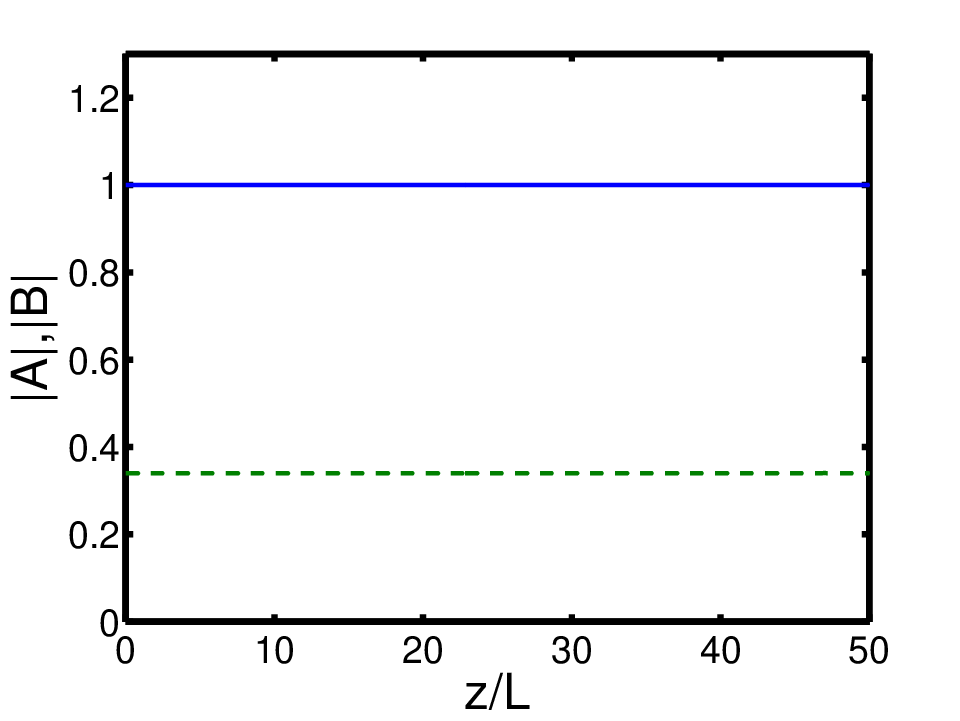}&\includegraphics[width=8cm]{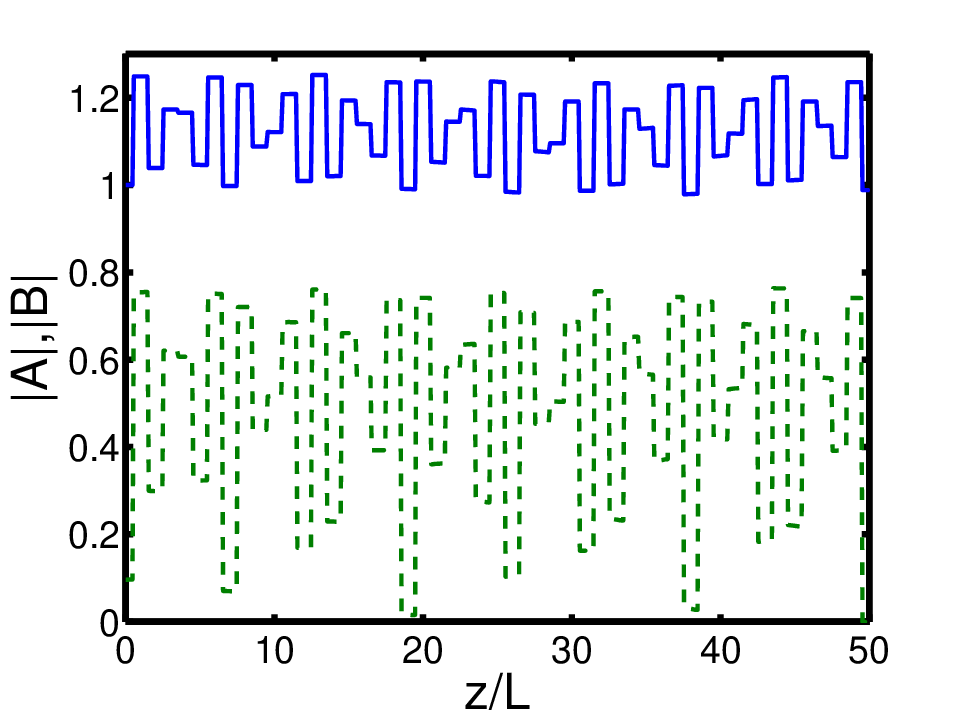}
\end{tabular}

\caption{Nonlinear propagative stationary solutions $N=50$, $\theta=0.8$, $\Gamma=0.0325$ and $\Omega=6.4271$. Solid line is the $A$ field and dashed the $B$ field. (a) Simulated non-reflecting boundaries and (b) physical finite-size structure.}\label{figu:modos_prop}
\end{figure}

{ We have considered that the solutions are CW signals with constant modulus when dealing with the effect of the nonlinearity in the previous derivation.  Therefore, the treatment of exponentially decaying signals in the forbidden band is not correct.  The results are approximately valid if we consider a case with $q_I N$ small so the intensity does not vary appreciably along the $N$ cells but, in general, we will over-estimate the impact of the nonlinearity. One could use the $n$-dependent nonlinear term to treat the effect of the variation of the signal intensity along the chain.  Instead, we will deal with this effect in a simpler though approximate manner.  To do so, we heuristically introduce an effective global nonlinearity and a corresponding effective wavenumber $q_{I,eff}$.  This is analogous to the previous definition of the effective distance $d_{eff}$ for the nonlinear contribution in the presence of losses: In a continuous system, the magnitude of the nonlinear phase shift can be obtained as $\gamma I(0) d_{eff}=\gamma \int_0^dI(z)dz$, with $I=\left|E\right|^2$. For our discrete system, we substitute the integral with a sum and similarly define, for a region under analysis spanning $N$ cells,
\begin{equation}
\Gamma_{eff}=\Gamma\frac{1}{N}\sum_{n=0}^{N-1}\exp\left(-2 q_{I,eff}n\right)=\frac{\Gamma\left(1-\exp\left(-2 q_{I,eff} N\right)\right)}{N\left(1-\exp\left(-2 q_{I,eff}\right)\right)},\label{ecu:factor}
\end{equation}
and a corresponding correction factor
\begin{equation}
g_{eff}=\frac{\Gamma_{eff}}{\Gamma}.
\end{equation}

In the limit $q_{I,eff}\to 0 $ one recovers $\Gamma_{eff}\to\Gamma$. Since $\Gamma$ and $q_I$ are mutually related, one requires the consistent calculation of both parameters.  This can be done by simply iterating Eqs. (\ref{ecu:factor}) and (\ref{eq:ate}).  We note that the effective nonlinearity parameter that approximates the global nonlinear contribution from a chain with fixed length $N$ is dependent on the value of $N$. 
}

{For finite length structures, boundary effects will result in the excitation of backward propagating modes.  Furthermore, the presence of counter-propagating waves has not been considered in the derivation of the exact nonlinear steady-state wave solution but will modify the propagation properties.}
{ We now study the properties of nonlinear steady-state solutions in unbounded reflectionless structures and in realistic finite-size resonator chains.  The former will permit to check the validity of the numerical algorithm whereas the latter will illustrate some of the interface effects will find in the following sections}. { It is convenient to substitute $P= |A|^2\left(1+\left|f\right|^2\right)$, which reduces to  $P=2 |A|^2$ for $|f|=1$ nonpropagative solutions, in Equations (\ref{eq:dispersion_nl}) to (\ref{eq:f}).  From the resulting expressions, we observe that commonly $\Gamma$ and $|A_1|^2$ are equivalent control parameters for this system.}

{
The profile of nonpropagative steady-state solutions, together with that of the corresponding theoretical predictions obtained from Eq. (\ref{eq:ate}), are displayed in Fig. \ref{figu:modos_ev} for the linear case.  We find an excellent match between the theory and simulations in non-reflecting structures, as shown in Fig.\ref{figu:modos_ev}.b.  For physical boundary conditions,  the $B$ field component is fixed to zero at the output and a perfect reflection of the incident signal is obtained.  The reflected signal arrives largely debilitated to the opposite end at $n=1$ where any further reflection becomes negligible.  Therefore, the input condition effectively fixes the value of the incident mode.  At the output end, the perfect reflection has the effect to rise the value of the $A$ component at the output plane, as illustrated in Fig. \ref{figu:modos_ev}, to a value close to twice the incident amplitude.  This is a situation typtical of perfect reflections in electromagnetic problems where the cancelation of the $B$ component will produce a reflected signal of the same amplitude as the incident such that their interference will produce a larger value for $A$.  This effect is of little significance if the field at the end of the structure, as in Fig. \ref{figu:modos_ev}.c, is already very small, but leads to a substantial distortion of the field distribution for chains of length not too large compared to $q_I^{-1}$, as shown in Fig. \ref{figu:modos_ev}.d.
}

{
Fig. \ref{figu:fit} compares the analytical and numerical results for nonlinear CW solutions in the forbidden band.  Very good corresponce between the numerical and the analytical results incorporating the effective nonlinearity correction (dashed-dotted line) is found.  Red dots indicate the uncorrected analytical solutions.  $g_{eff}=0.391$ for the short $N=10$ chain in the left pannel.  When the length of the structure is increased to $N=20$, the effective nonlinearity diminishes with $g_{eff}=0.203$.  If we increase the nonlinearity to $\Gamma=0.1$, the same structures give effective correction values of $g_{eff}=0.452$ and $g_{eff}=0.209$ for $N=10$ and $N=20$ rings, respectively, whereas for this value of $\Gamma$ the uncorrected solution predicts a fully propagative mode. 
 }

{For the study of the nonlinear propagative steady-state solution, we we now take the inverse cosine in Eq. (\ref{eq:dispersion_nl}) to obtain an explicit expression for $Q$ and substitute the value of $\Delta Q$ in Eq. (\ref{eq:condicion}), the numerical values of $Q$ and $|f|^2$ can be obtained by a simple fixed point iteration of the resulting relation together with Equation (\ref{eq:f}).  Starting with reasonable values for $|f|^2$ and $Q$, convergence typically follows after very few iterations.  Fig. \ref{figu:modos_prop} displays the results for a structure with $N=50$ and reflectionless (a) and a physical finite-size boundary (b) conditions.  The iterative calculation provides a value of $|f|\simeq 0.34$ in excellent agreement with the numerical results displayed in Fig. \ref{figu:modos_prop}.a.  When the effects of the reflections at the two boundaries of the structure are taken into account, after a transient, the steady-state standing wave pattern shown in Fig.\ref{figu:modos_prop}.b is obtained.}

\section{Optical switching in tightly coupled chains}

The nonlinearity has the effect of shifting the dispersion relation in the forbidden band along the $\Omega$ axis.  An input signal with a frequency inside a forbidden band will experience a transmission attenuation dependent on the propagation nonlinearity as the imaginary part of $Q$ is modified according to (\ref{eq:ate}).  This mechanism explains the switching behaviour of this type of structure \cite{broderick}. { Even though a single ring can be effectively used for switching, it is expected that a long distributed structure where a change in the wavenumber can have a large effect in the output, would permit a potential reduction in the switching power.  Nevertheless, the effective reduction of the nonlinear response in long chains described in the previous section indicates the possible existence of very steep switching responses in these structures with abrupt transitions from a quasi-linear regime.}

We have numerically addressed the switching effect by placing a unit amplitude input signal $\left|A_1\right|=1$ with a frequency $\Omega=3\pi/2$ (in the middle of the second forbidden band) for different values of $\Gamma$.  { We have also checked that} this is fully equivalent to keeping $\Gamma$ constant and changing the input intensity $|A_1|^2$ accordingly and that identical dynamics are produced when the input frequency is tuned to the middle of the first band $\Omega=\pi/2$.

\begin{figure}[ht]
\centering
\includegraphics[]{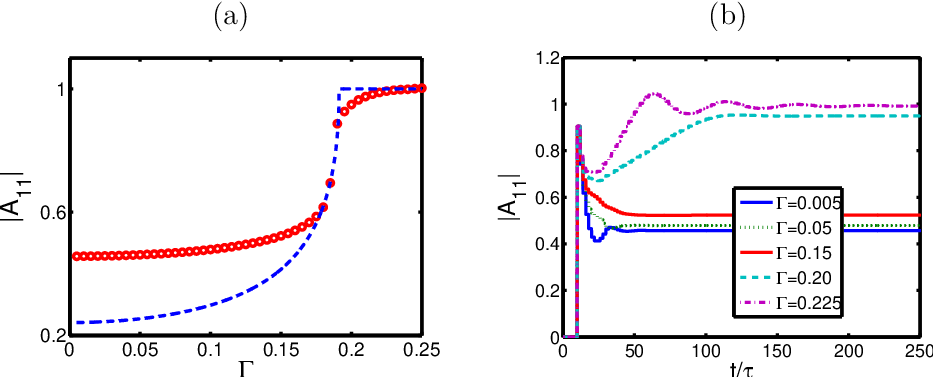}
\caption{(a) Nonlinear switching in a $N=10$ CROW with $\theta=0.99$ as obtained numerically (points) and fitted prediction from (\ref{eq:ate}) in the absence of boundary effects (dashed). (b) Transient response.}\label{figu:conmuta}
\end{figure}

In Fig.\ref{figu:conmuta}.a the numerical results of the transmission switching are displayed with points { whereas the dashed line corresponds to a qualitative fitting using our analytical framework}. {  The model presented in Section 2 does not encompass the full complexity of this nonlinear problem because of the presence of reflected nonlinear waves.   Still, the analytical results provide a reasonably good qualitative description of the switching behaviour.  For very small values of $\Gamma|A_1|^2$, in a quasi-linear regime, the numerical results in Fig. \ref{figu:conmuta}.a  show that the total output amplitude is nearly twice as large as $|A_{N+1}|$ as it would be expected from a perfect reflection of the attenuated linear mode.  This scenario was discussed in Section 3.  An arbitrary fit of the stationary solutions permits also to describe qualitatively the switching behaviour in the nonlinear regime.  The results for a forward propagating mode with $A_{N+1}=A_{1}\exp\left(-jQN\right)$ are shown with a dashed line, where $Q$ is obtained from Equation \ref{eq:ate} fitted with a particular value of $P$.    }  

   Fig. \ref{figu:conmuta}.b shows the output transient for various values of $\Gamma$. {A step input signal has been employed in the simulations.  As the system nonlinearity (or, equivalently, the input power) is raised, the transient response displays an increasing oscillatory behaviour.  For even larger values, the onset of a pulsating regime follows, similarly to the behaviour described in \cite{grigoriev2011}}.

\begin{figure}[ht]
\centering
\includegraphics[]{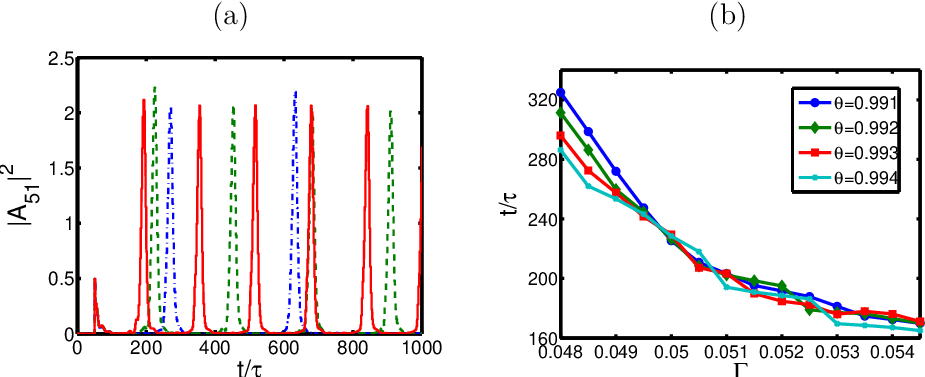}
\caption{(a) Spontaneously generated pulse trains for $\theta=0.993$ and $\Gamma |A_1|^2=0.047$ (dashed-dotted), $\Gamma |A_1|^2=0.050$ (dashed) and $\Gamma|A_1|^2 =0.055$ (solid). (b) Average pulse repetition period for varying $\Gamma |A_1|^2$ and four values of $\theta$.}\label{figu:solitones}
\end{figure}

{
\begin{figure}[ht]
\centering
\begin{tabular}{cc}
(a)&(b)\\
\includegraphics[width=8cm]{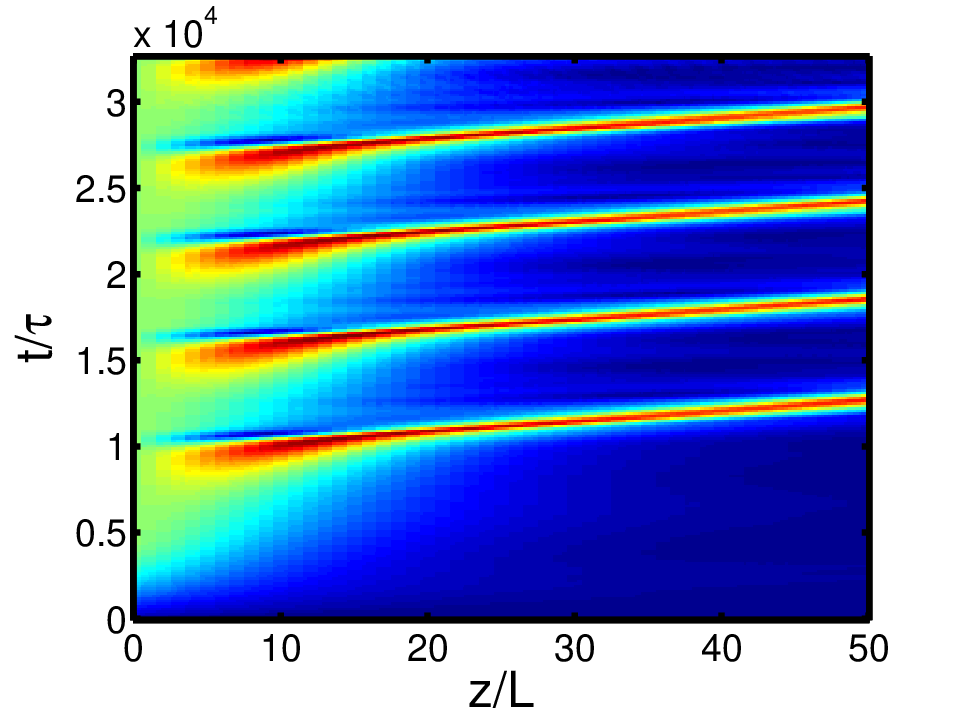}&\includegraphics[width=8cm]{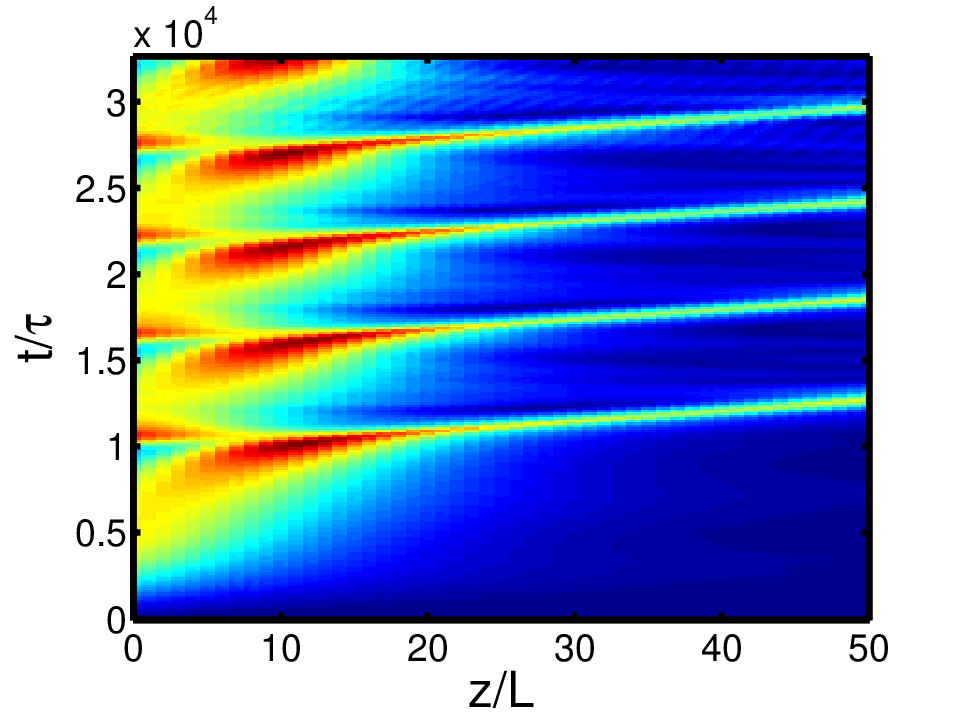}\\
(c)&(d)\\
\includegraphics[width=8cm]{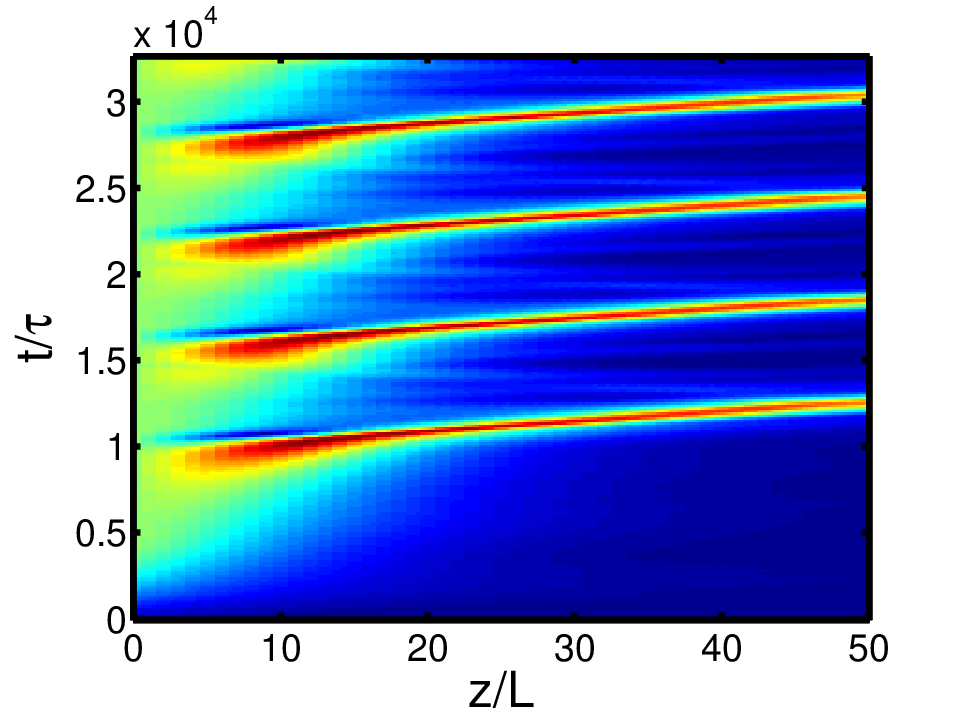}&\includegraphics[width=8cm]{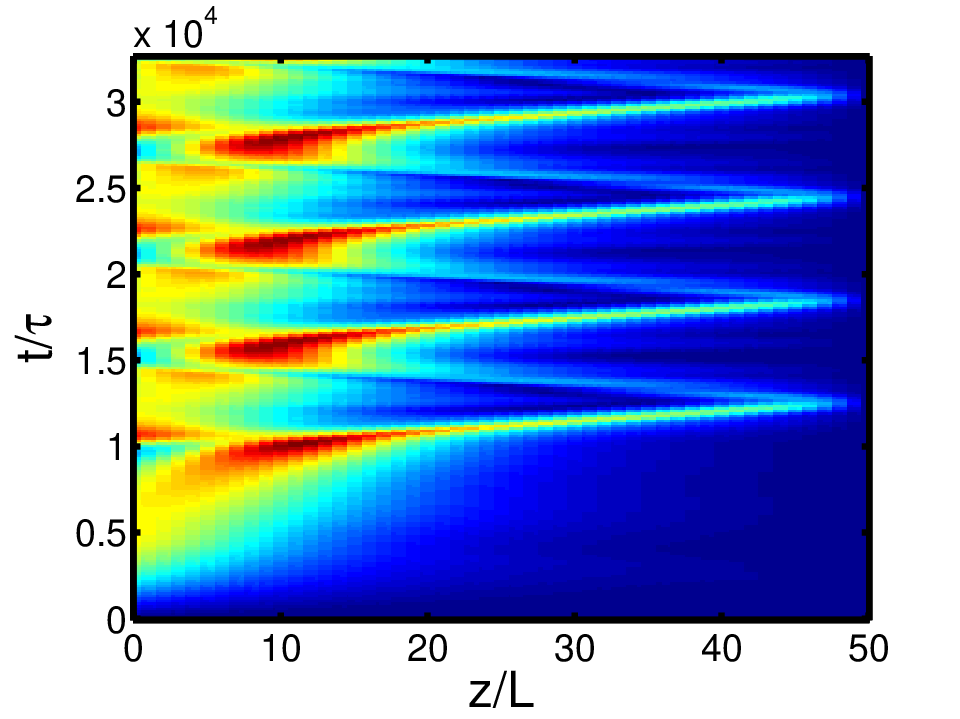}
\end{tabular}
\caption{Intra-cavity field amplitude dynamics with  for $\theta=0.993$, $|A_1|=1$ and $\Gamma=0.055$. (a) and (b) are the  $A$ and $B$ components, respectively, of an unbounded chain and (c) and (d) are the  $A$ and $B$ components, respectively, of an $N=50$ finite-size chain.}\label{figu:sol_dentro}
\end{figure}
}

\section{Spontaneous generation of Bragg solitons}

Inside the forbidden band, close the band edge, the dynamics are known to be well-described by a nonlinear Schr\"odinger equation \cite{neokosmidis,eggleton}.  It is anticipated that the nonlinear CW solutions are modulationally unstable for sufficiently high nonlinearity and that Bragg solitons \cite{christodoulides,aceves} arise as attractors of the nonlinear dynamics.  We address numerically this effect for a chain with $N=50$ cells in the tight-coupling regime $\theta\to 1$ by tuning the input frequency to a ten percent of the total frequency gap inside the forbidden band in each case.  Fig. \ref{figu:solitones}.a illustrates typical output pulse sequences obtained after the onset of the instability for $|A_1|=1$.  For these simulations, the intra-cavity field displays the expected generation and subsequent propagation of {non-dispersive pulses} inside the CROW.  Also, one can observe in Fig. \ref{figu:solitones}.a how the pulse spacing decreases as the nonlinearity increases.  This effect is illustrated in Fig. \ref{figu:solitones}.b for four different values of $\theta$ and values of $\Gamma$ in the range between  $0.048$ and $0.054$.

\begin{figure}[ht]
\centering
\begin{tabular}{cc}
(a)&(b)\\
\includegraphics[width=8cm]{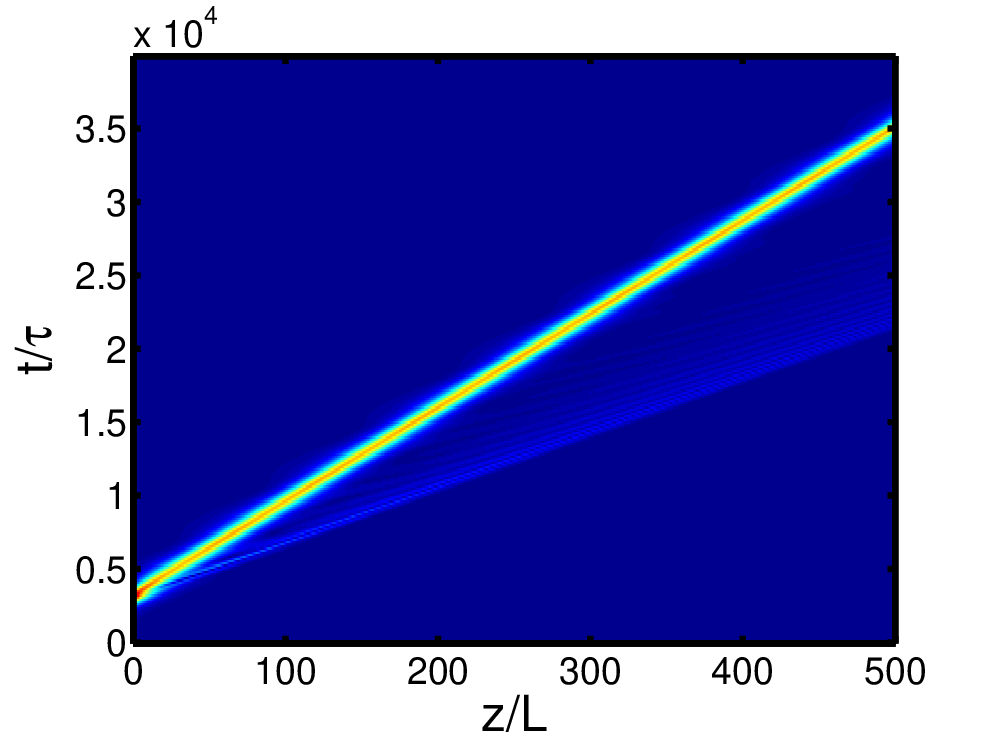}&\includegraphics[width=8cm]{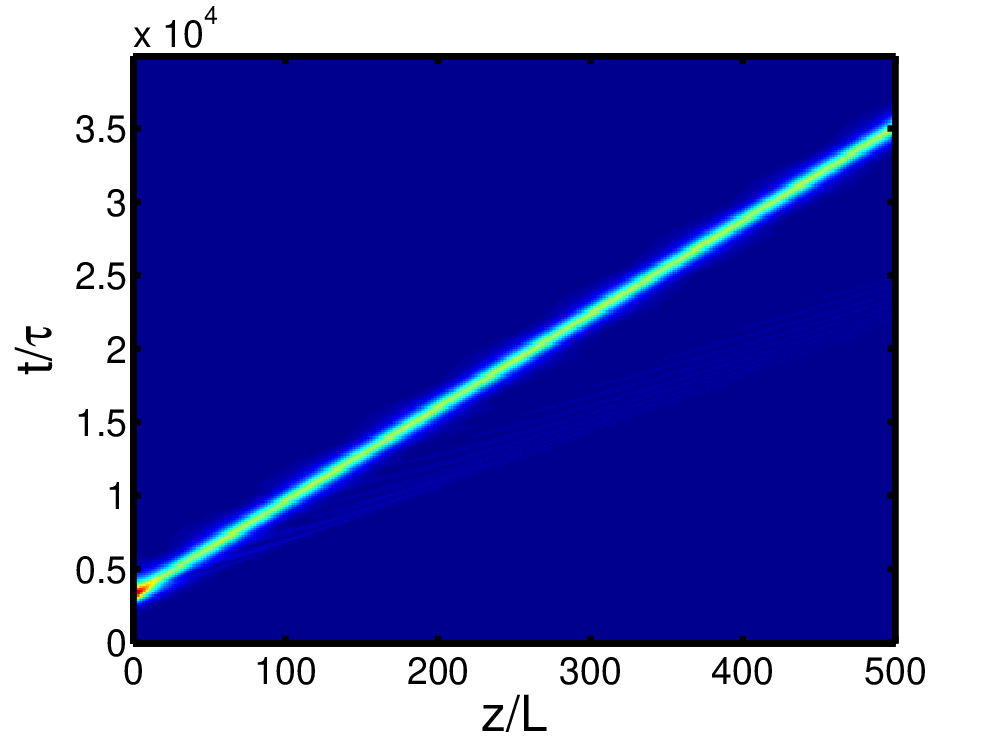}
\end{tabular}
\caption{Intra-cavity field dynamics in a long chain when a sech pulse with similar parameters to those of the output pulses in Fig. 7 c is re-injected at the input.   (a) and (b) are the  $A$ and $B$ components, respectively, and $\theta=0.993$ and $\Gamma=0.055$, as in Fig. 7.}\label{figu:sol_comprobacion}
\end{figure}

{ The details of the intracavity field components are displayed in Fig. \ref{figu:sol_dentro}.  Pannels (a) and (b) show the results for the simulation of a non-reflecting boundary (or, equivalently, an unbounded chain) whereas (c) and (d) correspond to a physical finite-size structure.  For these plots, a smoothly rising pulse instead of a step signal has been used.  One can clearly identify the effect of the reflections on the $B$ component, whereas the $A$ components of the field, which is largely responsible for the observed outputs, is very similar for the two cases.  Even though the role of this convective process associated to the back-and-forth movement of a localised nonlinear excitation in the pulsating behaviour is linked to the generation of gap solitons \cite{maes2009,sterke90} in finite structures, the numerical experiment we have conducted, namely, the generation of a pulsed output from a CW signal, corresponds precisely to the phenomenon of modulational instability.  This effect was studied formerly in side-coupled integrated spaced sequences of resonators (SCISSOR) \cite{heebner2002} using a nonlinear Schr\"odinger equation model and in coupled resonator waveguides of the type addressed here using an extended discrete nonlinear Schr\"odinger equation \cite{huang2009} obtained by means of the tight binding theory.}

{ We now elucidate whether the localised field structures found in the numerical simulations correspond to true Bragg solitons.  We point out that in finite chains, as commented above, only significant reflection is found for the $B$ component of the field, which is also attenuated upon reflection.  Therefore, the aforementioned back-and-forth movement cannot be associated with the full reflection of a gap soliton.  The output pulses generated from a CW input in the numerical experiments have a central portion that fits very well to a sech pulse which is surrounded by a smaller amplitude pedestal.  The complex envelope of the central part of the pulse displays a nearly constant phase, characteristic of soliton-like solutions.  We now reinject as $A_1$ a sech-shaped pulse with parameters close to those of the central portion of one of the output pulses at $A_N$ in the finite-length chain and study its evolution in a long coupled-microring structure.  The magnitudes of the propagating $A$ and $B$ components of the field are displayed in Figure \ref{figu:sol_comprobacion}.  After a short initial transient, a nondispersive pulse with the characteristics typically associated to a solitary wave propagates in the structure.  The initial transient is mainly due to the deviation from the exact solution in the $B$ field component intial conditions, since only the value of $A$ is fixed at the input.  During this transient, a small-amplitude dispersive wave is produced which propagates ahead of the slower soliton structure.  Both the existence of a transient period and the generation of a dispersive wave are typically associated to the generation of solitons from non-exact input conditions in a broad range of scenarios.} 

\section{Conclusion}

We have presented a study of the nonlinear wave propagation phenomena displayed by a nonlinear CROW.  A model for the system dynamics has been introduced and its exact nonlinear steady state modes for unbounded structures have been derived.  {The exact solutions of the model have been used to test the numerical scheme employed, having found a very good correspondence with the analytical predictions.  The properties of non-propagative solutions have to be corrected taking into account the decrease of the signal intensity along the CROW.  This has the result that nonpropagative stationary solutions in very long chains tend to behave quasi-linearly.  Nevertheless, complex dynamic behaviour can be found in this regime by the excitation of local perturbations.}  We have numerically studied the switching behaviour, which can be qualitatively explained using the stationary solutions, and the spontaneous generation of Bragg solitons.

\ack

We would like to thank David Novoa for useful discussions.  This work has been supported by Junta de Castilla y Le\'on, project VA001A08, and the Spanish Ministerio de Educaci\'on y Ciencia and Fondo Europeo de Desarrollo Regional, project TEC2010-21303-C04-04.

\section*{References}

\bibliography{refs}{}
\bibliographystyle{unsrt}

\end{document}